\documentclass[letterpaper,10pt]{revtex4}
\usepackage{graphicx}
\begin{document}

\title{A spaceborne gravity gradiometer concept based on cold atom interferometers for measuring Earth's gravity field}
\author{Olivier Carraz, Christian Siemes, Luca Massotti, Roger Haagmans, Pierluigi Silvestrin}
\affiliation{Earth Observation Programmes - European Space Agency \\
ESTEC P.O. Box 299, 2200 AG Noordwijk, The Netherlands \\
E-mail: olivier.carraz@esa.int - Fax: +31 (0) 71 565 4696}

\begin{abstract}
We propose a concept for future space gravity missions using cold atom interferometers for measuring the diagonal elements of the gravity gradient tensor and the spacecraft angular velocity. The aim is to achieve better performance than previous space gravity missions due to a very low white noise spectral behavior and a very high common mode rejection, with the ultimate goals of determining the fine structures of the gravity field with higher accuracy than GOCE and detecting time-variable signals in the gravity field better than GRACE.
\end{abstract}

\maketitle
\section{Introduction}
\label{intro}
Launched in 2002, the Gravity Recovery and Climate Experiment (GRACE) mission \cite{Tapley} measures changes in the EarthÕs gravity field, exploiting low-low satellite-to-satellite tracking (SST) using microwave K/Ka-band ranging and, like for CHAMP launched in 2000 \cite{Reigber}, high-low SST using GPS. The Gravity field and steady-state Ocean Circulation Explorer (GOCE) satellite \cite{Floberghagen}, launched in 2009, exploits high-low SST using GPS and, for the first time, gravity gradiometry. 
Continuing these successful gravity mission series, GRACE Follow-On (launch planned in 2017) will exploit low-low SST using K/Ka-band ranging and, as a demonstrator, laser ranging as well as high-low SST \cite{Sheard}. Future mission concepts like GRACE 2 or ESAÕs Next Generation Gravity Mission focus on low-low SST using laser ranging \cite{Silvestrin}. Electrostatic gravity gradiometers approach their ultimate performances, even if some improvement could be realized \cite{Zhu}.

In the past decades, it has been shown that atomic quantum sensors have the potential to drastically increase the performance of inertial measurements \cite{{Peters},{Sorrentino_gradio}}. These inertial sensors present a very low and spectrally white noise as opposed to classical accelerometers, which present colored noise. Recent results in different labs show that it is possible to build a reliable system for using atom interferometry on vehicles \cite{{Wu},{Bidel}} and for space applications \cite{{Yu},{Sorrentino_space},{Geiger}}. These last developments prove that this technology, which is already suitable on ground vehicles, is competitive with classic inertial sensors. Some developments already worked in zero-g environment in the drop tower facility in Bremen, Germany \cite{Muntinga}, or in a 0g plane \cite{Geiger} and are the state-of-the-art of compact setups aiming for spaceborne platform. Depending on the application current developments are now limited to a single component measurement \cite{Sorrentino_gradio}, or limited to a short interaction time due to gravity field on the ground \cite{Jiang}, or dedicated to fundamental physics such as Weak Equivalence Principle \cite{{Sorrentino_space},{Bonin}}  or the detection of gravitational waves \cite{Barrett}. 

We propose here a concept using cold atom interferometers for measuring all diagonal elements of the gravity gradient tensor and the full spacecraft angular velocity in order to achieve better performance than the GOCE gradiometer over a larger part of the spectrum, with the ultimate goals of determining the fine structures and time-variable signals in the gravity field better than today. This concept relies on a high common mode rejection \cite{Bonin} and a longer interaction time due to micro gravity environment, which will a provide better performance than any other cold atom interferometer on the ground, as already shown by a spectacular amelioration of the sensitivity using high fountain \cite{Dickerson}.

\section{Principle of an Atom Interferometer}
\label{sec1}
Although the wave nature of matter has been known since de Broglie, practical applications of matter waves have not been forthcoming until the appropriate techniques were devised and combined to cool, trap and manipulate atom ensembles and atom beams. These techniques have been continuously refined over nearly three decades resulting in the field of (cold) atom interferometry (AI) and the development of a host of laboratory instruments operating as inertial sensors or used for fundamental physics experiments.

Environmental vibrations, gravity fluctuations, and the presence of gravity Ñ effects that curb the interferometer interaction time Ñ currently limit the performance of atom interferometers on the ground. The operation of an atom interferometer in space under suitable conditions, i.e. in microgravity environment, will enable hitherto unachievable performance and allow for ultra precise measurements not possible on the ground. At low altitude residual drag can still exist but provided it is compensated to less than few mN this should not affect the measurements as the vibrations are rejected.

Atom interferometry will enable the realization of inertial and gravity sensing payloads with unprecedented sensitivity. Future Earth gravity missions will require gradiometers with sensitivity in the order of $1\;mE/\sqrt{Hz}$ ($1\;E = 10^{-9}\;s^{-2}$) over a wide spectral range. Atom interferometers with a baseline in the range of 1 meter are expected to meet this requirement and can be used for the realization of next-generation gravity gradiometer payloads \cite{Silvestrin}. The main advantages of this technology are a flat noise power spectral density even for low frequency measurements with very good repeatability, no hard moving parts and an intrinsically accurate measurement thanks to the stability of the atom transitions.

\begin{figure}
 \includegraphics[width=0.40\textwidth]{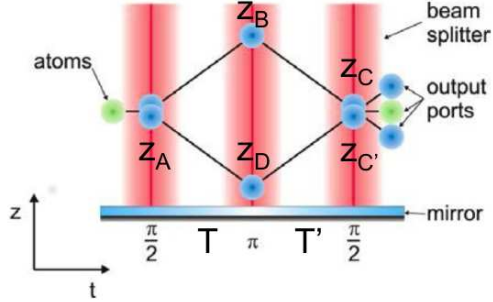}
   \caption{(Color online) The double diffraction scheme for one cloud of atoms. Figure from \cite{Tino}.} 
 \label{Double_diffraction}  
 \end{figure}

Atom interferometers rely on the wave-particle duality, which allows matter waves to interfere, and on the superposition principle. They can be sensitive to inertial forces. We present here the principle of a Chu-Bord\'e interferometer with $^{87}Rb$, which can be extended to any kind of inertial sensors such as a gravimeter \cite{Peters}, a gyroscope \cite{Canuel} or a gravity gradiometer \cite{Sorrentino_gradio}.
In a Chu-Bord\'e interferometer the test mass is a cloud of cold atoms, which is obtained from a Magneto-Optical Trap (MOT) by laser cooling and trapping techniques. This cloud of cold atoms is released from the trap and its acceleration due to external forces is measured by an atom interferometry technique. A Chu-Bord\'e interferometer consists in a sequence of three equally spaced Raman laser pulses \cite{Kasevich}, which drive stimulated Raman transitions between two stable states of the atoms. In the end, the proportion of atoms in the two stable states depends sinusoidally on the phase of the interferometer $\Phi$, which is proportional to the acceleration of the atoms along the Raman laser direction of propagation.
The Chu-Bord\'e interferometer in a double diffraction scheme (see Fig. \ref{Double_diffraction}) allows to enlarge the sensitivity by a factor 2, and to suppress at first order parasitic effects such as light shift, magnetic field, as the atoms remain in the same internal state \cite{{Leveque},{Giese}}. Phase shifts of this effects will be detailed in section \ref{sec3}.

In presence of an acceleration $a=a_{M}+ \gamma (z-z_{M})$ the trajectories of the atoms can be calculated using the Euler-Lagrange equation. 
In micro-gravity, $a_{M}$ is the acceleration experienced by the inertial reference, which is the mirror in Figure \ref{Double_diffraction}. $z$ and $z_{M}$ are the position of the atom cloud and the mirror, respectively, in the instrument frame. We decide by convention $z_{M}=0$. The gradient force is  $\gamma=V_{zz}-\Omega^{2}$, where $V_{zz}$ is the diagonal element of the gravity gradient tensor along z-axis defined by the Raman laser, and $\Omega^{2}=\omega_{x}^{2}+\omega_{y}^{2}$ is the rotation rate of the instrument. The initial position of the atom is $z_{A}$, with an initial velocity $v_{0}$ along z. A Raman pulse (called $\pi/2$ Raman pulse) of duration $\tau_s=\frac{\pi}{\Omega_{eff}\sqrt{2}}$, where $\Omega_{eff}$ is the Rabi effective frequency, produces a coherent superposition between two momentum states. Due to the Raman interaction, the clouds of atoms receive a 2 photons recoil velocity in the opposite direction. $\frac{\hbar k_{eff}}{2m}$ is the 2 photons recoil velocity due to Raman interaction \cite{Leveque}, where $m$ is the mass of the atom and $k_{eff}\approx\frac{8\pi}{\lambda}$ is the effective wave vector of the Raman transition. After free falling during a time $T$ the atoms are positioned at $z_{B}$ and $z_{D}$. A 4 photons recoil velocity in opposite direction via a so-called $\pi$ pulse (of duration $\tau_m=\sqrt{2}\pi/\Omega_{eff}$) is then given and after a time $T'$ a last $\pi/2$ pulse is given to the atoms to recombine the atom interferometers at positions $z_{C}$ and $z_{C'}$ (see Fig. \ref{Double_diffraction}).
\begin{equation}
\begin{array}{l l}
z_{A}=z_{0} \\
\begin{array}{l l}z_{B}=&\frac{a_{M}}{\gamma}+\left( z_{0}-\frac{a_{M}}{\gamma}\right)\cdot ch\left(\sqrt{\gamma}T\right) \\&+\left(\frac{v_{0}}{\sqrt{\gamma}}+\frac{\hbar k_{eff}}{2m\sqrt{\gamma}}\right)\cdot sh\left(\sqrt{\gamma}T\right) \end{array}\\
\begin{array}{l l}z_{D}=&\frac{a_{M}}{\gamma}+\left( z_{0}-\frac{a_{M}}{\gamma}\right)\cdot ch\left(\sqrt{\gamma}T\right)\\&+\left(\frac{v_{0}}{\sqrt{\gamma}}-\frac{\hbar k_{eff}}{2m\sqrt{\gamma}}\right)\cdot sh\left(\sqrt{\gamma}T\right)  \end{array}\\
\begin{array}{l l}z_{C}=&\frac{a_{M}}{\gamma}+\left( z_{0}-\frac{a_{M}}{\gamma}\right)\cdot ch\left(\sqrt{\gamma}\left(T+T'\right)\right)\\&+\left(\frac{v_{0}}{\sqrt{\gamma}}+\frac{\hbar k_{eff}}{2m\sqrt{\gamma}}\right)\cdot sh\left(\sqrt{\gamma}\left(T+T'\right)\right)\\&-\frac{\hbar k_{eff}}{m\sqrt{\gamma}}sh\left(\sqrt{\gamma}T'\right)\end{array}\\
\begin{array}{l l}z_{C'}=&\frac{a_{M}}{\gamma}+\left( z_{0}-\frac{a_{M}}{\gamma}\right)\cdot ch\left(\sqrt{\gamma}\left(T+T'\right)\right)\\&+\left(\frac{v_{0}}{\sqrt{\gamma}}-\frac{\hbar k_{eff}}{2m\sqrt{\gamma}}\right)\cdot sh\left(\sqrt{\gamma}\left(T+T'\right)\right)\\&+\frac{\hbar k_{eff}}{m\sqrt{\gamma}}sh\left(\sqrt{\gamma}T'\right)\end{array}
\end{array}
\end{equation}
The matter wave phase at the end of the interferometer is given by \cite{borde}:
 \begin{equation}
\begin{array}{l l}
\begin{array}{l l}\Phi &= \frac{\Phi_{ABDC}+\Phi_{ABDC'}}{2}\\&=k_{eff}\left(z_{C'}-z_{B}-z_{D}+z_{A}\right)+\frac{1}{2}k_{eff}\left(z_{C}-z_{C'}\right) \\
&=k_{eff}\left[\begin{array}{l l}\left( z_{0}-\frac{a_{M}}{\gamma}\right)\cdot \left(\begin{array}{l l}1+ch\left(\sqrt{\gamma}\left(T+T'\right)\right)\\-2ch\left(\sqrt{\gamma}T\right)\end{array}\right)\\+\frac{v_{0}}{\sqrt{\gamma}}\cdot \left(sh\left(\sqrt{\gamma}\left(T+T'\right)\right)-2sh\left(\sqrt{\gamma}T\right)\right)\end{array}\right] \\
&\approx k_{eff} \left[\left( z_{0}-\frac{a_{M}}{\gamma}\right)\left(\gamma T^{2}+\frac{7}{12}\gamma^{2}T^{4}\right)+v_{0}\gamma T^3\right]\end{array}
\end{array}
\end{equation}

\begin{figure}
\includegraphics[width=0.50\textwidth]{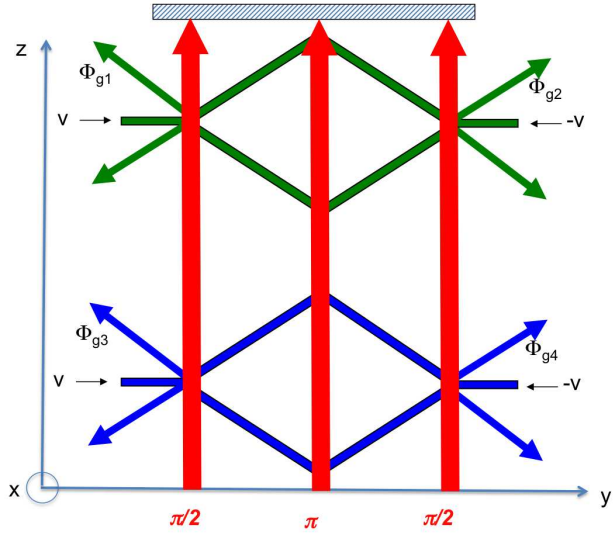}
   \caption{(Color online) Combining 4 atom interferometers to measure both gravity gradient and rotation rate.} 
 \label{Concept_interfero}  
 \end{figure}

It is possible to suppress $a_{M}$ by measuring simultaneously different atom interferometers \cite{{Sorrentino_gradio},{Dickerson}}. On Figure \ref{Concept_interfero} as the inertial reference being common to the 4 different atom interferometers, the high common rejection cancels the vibration noise and combining $\Phi g_{1}$ and $\Phi g_{2}$ (or $\Phi g_{3}$ and $\Phi g_{4}$) we can precisely measure the rotation rate $\omega_{x}$, knowing precisely $\vec{v}$ along y-axis. This is due to the measurement of the Coriolis effect $2\vec{\Omega}\times\vec{v}$ along z-axis. Combining  $\Phi g_{1}$ and $\Phi g_{3}$ (or $\Phi g_{2}$ and $\Phi g_{4}$) allows to measure $\gamma$, relying on the precise knowledge of the distance between the atom clouds. Applying this measurements in the 3 orthogonal directions (see Section \ref{subsec23}), the full angular velocity $\vec{\Omega}$ is simultaneously measured, thus we have direct access to all diagonal elements of the gravity gradient tensor $V_{xx},V_{yy},V_{zz}$.
 
Due to force gradient, the interferometer is not perfectly closed. The distance at the end of the interferometer is proportional to the 4 photons recoil velocity $4v_{rec}=\frac{\hbar k_{eff}}{m}$:
\begin{equation}
\begin{array}{l l}\Delta z&=z_{C}-z_{C'}\\ &=\frac{\hbar k_{eff}}{m\sqrt{\gamma}}\cdot\left[sh\left(\sqrt{\gamma}\left(T+T'\right)\right)-2sh\left(\sqrt{\gamma}T'\right)\right]\\&\approx 4v_{rec}\gamma T^{3}
\end{array}
\end{equation}
As in optical interferometers we introduce a coherence length of the atoms, which is defined by the wavelength of De Broglie, directly dependent on the temperature $\Theta$ of the atoms.
\begin{equation}
\lambda_{DB}=\frac{h}{\sqrt{2\pi mk_{B}\Theta}}
\end{equation}
For example Table \ref{tab_temp} reports the different coherence lengths needed, in relation to the temperature.
\begin{table}
\caption{Type of cold atom clouds needed for different interrogation time.}
\label{tab_temp}       
\begin{tabular}{llll}
\hline\noalign{\smallskip}
\textbf{T} & $\mathbf{\Delta z}$ & $\mathbf{\Theta}$ \textbf{for}  & \textbf{Cold technique} \\
& & $\mathbf{\Delta z = \frac{\lambda_{DB}}{2}}$ & \\
\noalign{\smallskip}\hline\noalign{\smallskip}
$1\;s$ & $36\;nm$ & $6.5\;\mu K$ & Molasses \\
\noalign{\smallskip}\hline\noalign{\smallskip}
$2\;s$ & $290\;nm$ & $100\;nK$ & Kinetic selection \\
\noalign{\smallskip}\hline\noalign{\smallskip}
$5\;s$ & $4.5\;\mu m$ & $0.4\;nK$ & Bose Einstein \\
 & & & Condensate (BEC) \\
\noalign{\smallskip}\hline
\end{tabular}
\end{table}

A Bose Einstein Condensate (BEC) is required for T= 5 s. Even if it is possible to close the interferometer by changing $T'$ \cite{Roura}, the maximum temperature estimated by the extension of the cloud of the atoms, has to be lower than 10 nK which corresponds to a cloud extension of 1 mm/s.
To avoid loss of contrast due to the initial velocity distribution of the atom source, a point source interferometry (PSI) scheme \cite{Dickerson} should be used to simultaneously detect the differential acceleration of  2 clouds of atoms.
As the scheme relies on differential acceleration measurements between two correlated, simultaneously operated atom interferometers (separated by a baseline d), then the error per measurement on the gravity gradient is:
\begin{equation}
\sigma_{\gamma}=\sigma_{\Phi}\cdot\frac{1}{k_{eff}}\cdot\frac{1}{T^{2}}\cdot\frac{\sqrt{2}}{d}
 \label{eq_err}
 \end{equation}
The error on the phase is $\sigma_{\Phi}=\frac{1}{SNR}$. The signal to noise ratio (SNR) depends on the detection system and the standard quantum limit (SQL). It is possible to reject the detection noise by a very low noise detection system \cite{Biedermann}, but some time is necessary until the interferometer output ports are separated for simultaneous detection. For a separation of 2.5 cm, 1 s is needed before detection. Limited by the SQL means that the SNR is proportional to the square root of the number N of atoms in the interferometer. With typically $N=10^{6}$ atoms, it is reasonable to assume a phase noise due to the detection at the level of 1 mrad as current state of the art.

\section{Concept of a spaceborne gravity gradiometer based on laser-cooled atom interferometry}
 \label{sec2}
  \begin{figure*}
\includegraphics[width=0.75\textwidth]{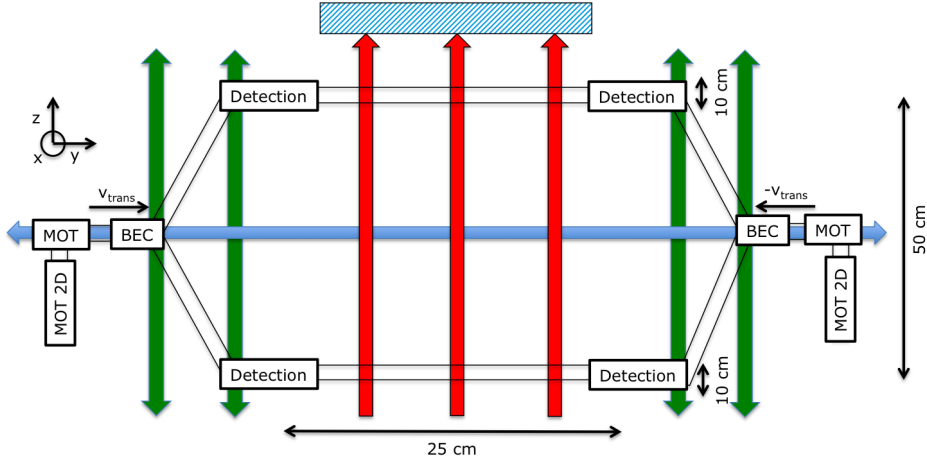}
   \caption{(Color online) Scheme of the vacuum chamber (not drawn to scale; orientative sizing). The red arrows represent the Raman lasers for the interferometry part; the blue arrows represent the light pulse giving the transverse velocity $v_{trans}$; the green arrows represent the light pulses guiding the atoms from the cooling room to the interferometer room; the blue rectangle is the mirror, which will be the inertial reference.} 
 \label{Concept_system}  
 \end{figure*}
 
Figure \ref{Concept_system} describes the gravity gradiometer concept in one dimension. This one-dimensional concept consists in measuring one diagonal element of the gravity tensor ($V_{zz}$) and the rotation rate along the x-axis. It can be extended to the other two dimensions in order to obtain the full diagonal elements of the gravity gradient tensor and the full angular rate vector.

\subsection{Preparation of the 4 cold atoms test masses}
\label{subsec21}
Two clouds of atoms are symmetrically prepared along the y-axis. In order to reach ultra-high vacuum inside the vacuum chamber and the high atom number required for efficient evaporative cooling, the MOT is loaded from a 2D-MOT flux \cite{Schoser}. The atoms are launched along y-axis with a precise velocity $\left|\left|\vec{v}\right|\right|=v_{trans}$ corresponding to 2 cycling transitions resulting in a 4 photons recoil velocity ($v_{trans}=4.v_{rec} = 2.5\;cm/s$). The laser providing this recoil moment is represented by a blue arrow in Figure \ref{Concept_system} and is common to the 2 MOTs. This will define the longitudinal axis for the rotation measurement. As the cloud of atoms is moving along y-direction, it is possible to load a new MOT for the following measurement while preparing the BEC: the moving atoms are first stopped by an opposite 2 cycling transition, then the BEC is prepared before first mentioned recoil laser is used to simultaneously move the second MOT in preparation and the BEC. Then a high recoil laser pulse \cite{Chiow} splits each cloud of atom in two and brings the 4 clouds to the interferometer rooms (green arrows in Fig. \ref{Concept_system}) along the z-axis. This defines the gradient axis. A second high recoil pulse brings them back to a parallel launch. With a 100 recoils velocity pulse it is thus possible to bring the atoms in the 50 cm separated interferometer rooms in 0.8 s. As the interferometer time can be assumed to be 2T = 10 s, the distance between two counter-propagating BEC is 25 cm in the interferometer sequence.

\subsection{Interferometer sequence}
\label{subsec22}
The Chu-Bord\'e interferometer is realized by using a double diffraction scheme with two counter-propagating Raman lasers (red arrows in Figure \ref{Concept_system}). It is thus possible to have a gradiometer by measuring the acceleration of two clouds of cold atoms from two MOTs separated by d = 50 cm. The use of counter-propagating atomic clouds allows measurements of the rotation rate along the x-axis (rotation axis).

As the interferometer is then delocalized it is also possible to prepare cold atoms during the interferometric sequence and then reduce the cycling time below the time of the interferometer measurement. It is thus possible to have a quasi-continuous measurement by launching atoms at a high frequency, limited by the time $T_{cycle}$ to produce cold atoms. Thus the interferometer will produce measurements for $f<1/T_{cycle}$. Best performances reported up to date are a BEC of $10^4$ atoms loaded in $1\;s$ \cite{Zoest}. We can expect for future experiments a BEC of few $10^6$ atoms loaded in $T_{cycle} = 1\;s$ \cite{Muntinga}.

According to Equation \ref{eq_err} and considering that the interferometer noise is limited by the quantum projection noise with $N=10^{6}$ atoms, the sensitivity that can be reached is given by the following expression, for $f<1/(2T)$:
\begin{equation}
\begin{array}{l l}
\Delta\gamma=\frac{\sqrt{2}}{\sqrt{N}\cdot k_{eff}\cdot d\cdot T^{2}}\cdot \sqrt{T_{cycle}}=3.5\;mE/\sqrt{Hz} \\
\Delta\omega=\frac{1}{2\cdot \sqrt{N}\cdot k_{eff}\cdot v_{trans}\cdot T^{2}}\cdot \sqrt{T_{cycle}}=25\;prad.s^{-1}/\sqrt{Hz}
\end{array}
\end{equation}

\begin{figure}
\includegraphics[width=0.50\textwidth]{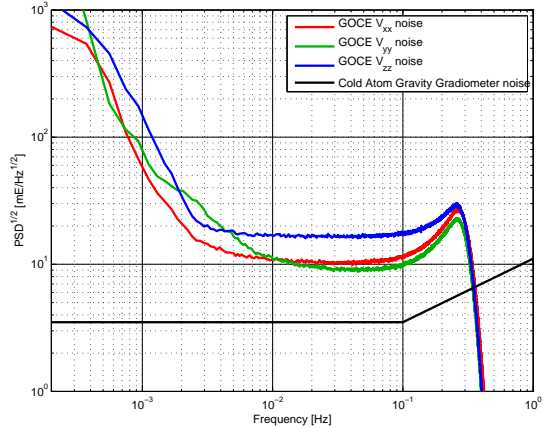}
   \caption{(Color online) Comparison between classical (GOCE)/quantum concepts for a space gradiometer.} 
 \label{PSD}  
 \end{figure}

As the time between each measurement is shorter than the interferometer time, for $1/(2T) < f < 1/T_{cycle}$ the sensitivity is increased by a factor $(2T.f)^{1/2}$ (see Fig. \ref{PSD}).

\subsection{Extension to the three directions}
\label{subsec23}
The scheme shown on Figure \ref{Concept_system} has to be duplicated in the 3 directions to have all diagonal elements of the gravity gradient tensor $V_{xx},V_{yy},V_{zz}$ and the angular rate vector $\vec{\Omega}$.

In order to avoid any overlap between the different laser pulses, the $\pi$ pulse in the middle of the interferometer can not be done anymore. Instead, having 4 Raman pulses for the interferometer sequence allows duplicating this one-dimensional concept in the three directions and then measuring the full diagonal elements of the gravity gradient tensor and the full angular rate vector.

\begin{figure*}
   \includegraphics[width=0.75\textwidth]{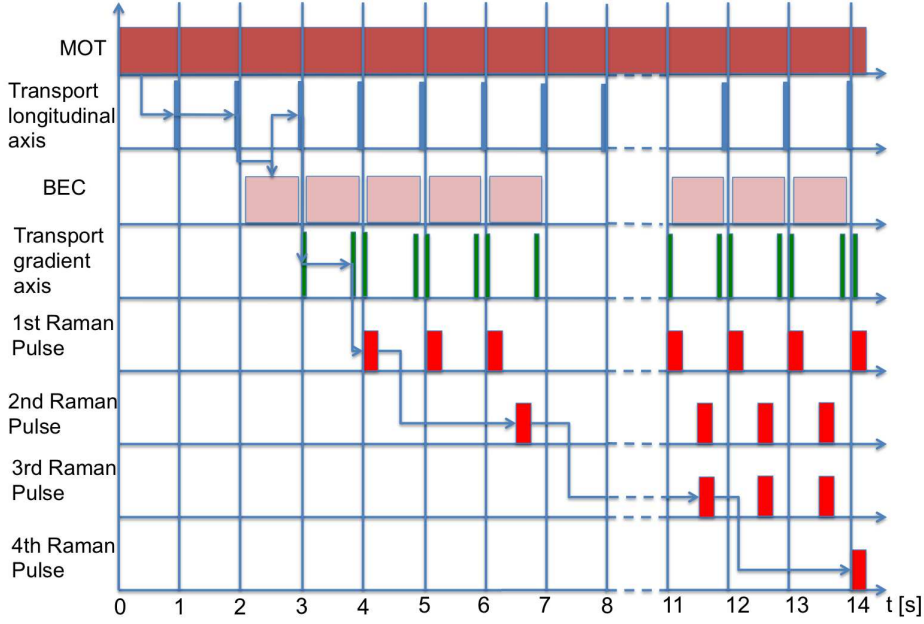}
   \caption{(Color online) Time sequence for measuring the diagonal elements of the gravity tensor and the rotation rates. The arrows indicate one point measurement.} 
 \label{Time_seq}  
 \end{figure*}

The height of each interferometer is $v_{rec}.T = 20\;cm$. This can be reduced by a factor 2 by implementing a Ramsey-Bord\'e interferometer with four pulses $\pi/2$ separated by T1 = 2.5 s, T2 = 5 s, T3 = T1 = 2.5 s, in which case the sensitivity is degraded by a factor 4/3, thus:
\begin{equation}
\begin{array}{l l}
\Delta\gamma=4.7\;mE/\sqrt{Hz} \\
\Delta\omega=35\;prad.s^{-1}/\sqrt{Hz}
\end{array}
\end{equation}
Due to losses at the second and third pulse, $4.10^6$ atoms have to be loaded before the interferometer to achieve this sensitivity. Moreover each arm should have a magnetic bias field in the direction of the Raman beams.
To avoid any overlap between the different laser functions the time sequences have to be very well controlled, as follows on Figure  \ref{Time_seq}.
 
\section{External noise sources}
\label{sec3}
Gravity measurement by atom interferometry is an absolute measurement, as the interferometric phase only relies on atomic transitions. Thus other phase contributions have to be very well determined to reach the desired high sensitivity. These phase contributions have different origins:
\begin{itemize}
\item Energy modification of the hyperfine structure of the Rubidium (Zeeman effect, AC Stark shift).
\item Axis of measurement (Wavefront curvature of the Raman laser, verticality of the axis).
\item Non-gravitational forces (Centrifugal acceleration, Coriolis force).
\end{itemize}
As the atoms are sensitive to the magnetic field, for a Chu-Bord\'e interferometer the dependence is proportional to the gradient of the magnetic field along the interferometer: $\delta\Phi=\alpha\cdot B_{0}\cdot\nabla B\cdot L\cdot T$ where $\alpha=2\pi.575\;Hz/G^{2}$ for $^{87}Rb\;m_{F}=0$, $B_{0}=1\;mG$ is the magnetic field applied during the interferometer sequence to avoid Zeeman transitions \cite{Peters} and L is the longitudinal length of the interferometer.

Due to matter-light interaction during Raman pulses, the AC Stark shift affects also the energy of the Rb. The first order phase shift is given by \cite{Peters}:
\begin{equation}
\Delta\Phi^{AC}=\frac{\delta_{4}^{AC}-\delta_{1}^{AC}}{\Omega_{eff}}
\label{eq_acshift}
\end{equation}
$\delta_{1}^{AC}$ (resp. $\delta_{4}^{AC}$) is the Stark shift during the first (resp. last) pulse.
With a well-defined ratio between the intensity of the Raman lines $I_{1}$ and $I_{2}$ it is thus possible to cancel this effect \cite{Peters}. 

Using the double diffraction scheme allows us to suppress these effects on the energy levels as the atoms remain in the same state, on condition that there is no magnetic field gradient and the Raman intensities do not fluctuate during the interferometer sequence.

Due to the expansion of the cloud of atoms, the wavevector defined by the Raman lines is not the same for different class of velocity of the atoms \cite{Louchet}, the error in acceleration on each cloud is at first order $\frac{\sigma_{v}^2}{R}$  where R is the wavefront curvature of the laser and $\sigma_{v}=1\;mm/s$ for a temperature of the cloud $\Theta=10\;nK$. For differential measurements, the error on the phase is $\delta\phi=k_{eff}T^{2}\sigma_{v}^2\frac{\delta R}{R^{2}}$ where $\delta R$ is the differential wavefront curvature of the lasers.

The axis of the Raman laser has to be well determined as the measurement is done along this axis. The relative error on the gravity gradient is half of the square of the angle error of the axis (attitude control of the laser direction), i.e. $\theta<2\sqrt{\Delta\gamma/\gamma}$.

Finally, using the interferometer as a gradiometer cancels common mode vibrations, and using it as a gyroscope leads to know the Coriolis force (considering $v_{\perp}<\sigma_{v}$), thus the centrifugal force that is a systematic effect on the gravity gradient measurement.

The table \ref{tab_error} gives the required maximum incertitude to reach on these different contributions on the interferometric phase, in order to get below a sensitivity of 1 mrad phase noise.
\begin{table}
\caption{External sources and minimum knowledge of these sources to achieve 1 mrad noise on the interferometric phase.}
\label{tab_error}       
\begin{tabular}{lll}
\hline\noalign{\smallskip}
\textbf{Noise source} & \textbf{Limit reached for 1 mrad} \\
\noalign{\smallskip}\hline\noalign{\smallskip}Ê
Magnetic field & $\frac{\delta B_{z}}{\delta y} < 0.1\;mG/m$ \\
\noalign{\smallskip}\hline\noalign{\smallskip}
AC Stark shift & $\frac{\delta\frac{I_{1}}{I_{2}}}{\frac{I_{1}}{I_{2}}}<0.1\%$ \\
\noalign{\smallskip}\hline\noalign{\smallskip}
Coriolis force & $\Omega_{\perp}<0.9\;nrad.s^{-1}$ \\
 & $v_{\perp}<1\;mm.s^{-1}$ \\
\noalign{\smallskip}\hline\noalign{\smallskip}
Centrifugal force & $\Omega<1.5\;\mu rad.s^{-1}$ \\
\noalign{\smallskip}\hline\noalign{\smallskip}
Wavefront curvature & $\frac{\delta R}{R^{2}}<1.5\times 10^{-6} \;m^{-1}$ \\
\noalign{\smallskip}\hline\noalign{\smallskip}
Raman axis & $\theta < 1.5\;mrad$ \\
\noalign{\smallskip}\hline
\end{tabular}
\end{table}

\section{Conclusion and outlook}
\label{concl}
A concept of gravity gradiometer based on cold atom interferometer techniques is proposed. This instrument allows reaching sensitivity of $3.5\;mE/\sqrt{Hz}$, with the promise of a flat noise power spectral density also at low frequency, and a very high accuracy on rotation rates. 
New techniques allow us to go beyond the SQL limit with squeezed state or Information-recycling beam splitters \cite{{Gross},{Haine}}. At best it is possible to reach the Heisenberg limit, i.e. to have a SNR proportional to N. The phase noise could be extended at best to $1\;\mu rad$ with these new techniques \cite{{Gross},{Haine}}, on condition than other noise sources can be consistent with this limit.

Estimation of the Earth gravity field model from the new gravity gradiometer concept has to be evaluated taking into account different system parameters such as attitude control, altitude of the satellite, time duration of the mission, etc., which is the subject of a separate analysis.

In the meantime as long as this gravity gradiometer concept is not yet available, hybridization between quantum and classical techniques could be an option to improve the performance of accelerometers on next generation gravity missions. This could be achieved as it is realized in frequency measurements where quartz oscillators are phase locked on atomic or optical clocks \cite{Vanier}. This technique could correct the spectrally colored noise of the electrostatic accelerometers in the lower frequencies.

\begin{acknowledgements}
The authors acknowledge useful discussions with Bruno Leone during the preparation of this manuscript.
The final publication is available at Springer via http://dx.doi.org/10.1007/s12217-014-9385-x
\end{acknowledgements}


\begin{thebibliography}{}
\bibitem{Tapley} B. Tapley et al., GRACE measurements of mass variability in the Earth system, Science (New York NY), 305(5683), 503Ð505 (2004)
\bibitem{Reigber} C. Reigber et al., The CHAMP geopotential mission, Bolletino di Geosica Teorica ed Applicata, 40, 285-289 (1999)
\bibitem{Floberghagen} R. Floberghagen et al., Mission design, operation and exploitation of the gravity field and steady-state ocean circulation explorer mission, Journal of Geodesy, 85, 749-758 (2011)
\bibitem{Sheard} B. Sheard et al., Intersatellite laser ranging instrument for the GRACE follow-on mission, Journal of Geodesy, 86(12), 1083-1095 (2012)
\bibitem{Silvestrin} P. Silvestrin et al., The Future of the Satellite Gravimetry After the GOCE Mission, International Association of Geodesy Symposia, Volume 136, 223-230 (2012)
\bibitem{Zhu} Z. Zhu et al., Electrostatic gravity gradiometer design for the future mission, Advances in Space Research, 51, 2269-2276 (2013)
\bibitem{Peters} A. Peters et al., High-precision gravity measurements using atom interferometry, Metrologia 38, 25-61 (2001)
\bibitem{Sorrentino_gradio} F. Sorrentino et al., Sensitivity limits of a Raman atom interferometer as a gravity gradiometer, Phys. Rev. A 89, 023607 (2014)
\bibitem{Wu} X. Wu, Gravity gradient survey with a mobile atom interferometer, Ph.D. dissertation, Stanford University (2009)
\bibitem{Bidel} Y. Bidel et al., Compact cold atom gravimeter for field applications, App. Phys. Let., 102 (2013)
\bibitem{Yu} N. Yu et al., Development of an atom-interferometer gravity gradiometer for gravity measurement from space, Appl. Phys. B, 84, 647 (2006)
\bibitem{Sorrentino_space} F. Sorrentino et al., A Compact Atom Interferometer for Future Space Missions, Microgravity Science and Technology, Volume 22, Issue 4, 551-561 (2010)
\bibitem{Geiger} R. Geiger et al., Detecting inertial effects with airborne matter-wave interferometry, Nature Communications 2, 474 (2011)
\bibitem{Muntinga} H. M\"untinga et al., Interferometry with Bose-Einstein Condensates in Microgravity, Phys. Rev. Lett. 110, 093602 (2013)
\bibitem{Jiang} Z. Jiang et al., On the gravimetric contribution to the redefinition of the kilogram, Metrologia 50, 452-471 (2013) 
\bibitem{Bonin} A. Bonin et al., Simultaneous dual-species matter-wave accelerometer, Phys. Rev. A 88, 043615 (2013)
\bibitem{Barrett} B. Barrett et al., Mobile and Remote Inertial Sensing with Atom Interferometers, Proceedings of the Enrico Fermi International School of Physics Enrico Fermi, Course 188, Varenna (2013)
\bibitem{Dickerson} S. M. Dickerson et al., Multiaxis Inertial Sensing with Long-Time Point Source Atom Interferometry, Phys. Rev. Lett. 111, 083001 (2013)
\bibitem{Canuel} B. Canuel et al., Six-Axis Inertial Sensor Using Cold-Atom Interferometry, Physical Review Letters 97, 010402 (2006)
\bibitem{Kasevich} M. A. Kasevich and S. Chu, Atomic interferometry using stimulated Raman transitions, Phys. Rev. Let. 67, 181-184 (1991)
\bibitem{Leveque} T. L\'ev\`eque et al., Enhancing the area of a Raman atom interferometer using a versatile double-diffraction technique, Physical Review Letters 103, 080405 (2009)
\bibitem{Giese} E. Giese et al., Double Bragg diffraction: A tool for atom optics, Phys. Rev. A 88, 053608 (2013)
\bibitem{Tino} G. M. Tino et al., Precision Gravity Tests with Atom Interferometry in Space, Nuclear Physics B (Proc. Suppl.), 243-244, 203-217 (2013)  
\bibitem{borde} C. J. Bord\'e, Theoretical tools for atom optics and interferometry, C.R. Acad. Sci. Paris, t.2, SŽrie IV, 509-530 (2001). 
\bibitem{Roura} A. Roura et al., Overcoming loss of contrast in atom interferometry due to gravity gradients, arXiv:1401.7699v1 (2014)
\bibitem{Biedermann} G. W. Biedermann et al., Low-noise simultaneous fluorescence detection of two atomic states, Optics Letters 34, 3 (2009)
\bibitem{Schoser} J. Schoser et al., Intense source of cold Rb atoms from a pure two-dimensional magneto-optical trap, Phys. Rev. A 66, 023410 (2002)
\bibitem{Chiow} S. W. Chiow et al., 102hk Large Area Atom Interferometers, PRL 107, 130403 (2011)
\bibitem{Zoest} T. van Zoest et al., Bose-Einstein Condensation in Microgravity, Science 328 (5985): 1540-1543 (2010)
\bibitem{Louchet} A. Louchet-Chauvet et al., The influence of transverse motion within an atomic gravimeter, New J. Phys. 13, 065025 (2011)
\bibitem{Gross} C. Gross, Spin squeezing, entanglement and quantum metrology with Bose-Einstein condensates, J. Phys. B: At. Mol. Opt. Phys. 45, 103001 (2012)
\bibitem{Haine} S. A. Haine, Information-recycling beam splitters for quantum enhanced atom interferometry, Physical Review Letters 110, 053002 (2013)
\bibitem{Vanier} J. Vanier, Transfer of Frequency Stability from an Atomic Frequency Reference to a Quartz-Crystal Oscillator, IEEE Transactions on Instrumentation and Measurement 28, 3 (1979)
\end{thebibliography}
\end{document}